\documentclass[aps,12pt,prd,preprintnumbers,showpacs,nofootinbib]{revtex4}

\usepackage{amsmath}
\usepackage{amsfonts}

\newcommand{\p}[1]{(\ref{#1})}
\newcommand{\lb}{\label}
\newcommand{\ca}[1]{{\cal #1}}

\newcommand{\be}{\begin{eqnarray}}
\newcommand{\ee}{\end{eqnarray}}

\newcommand{\D}{{\cal D}}
\newcommand{\F}{{\cal F}}
\newcommand{\A}{{\cal A}}
\newcommand{\B}{{\cal B}}

\newcommand{\SU}{{\rm SU}}
\newcommand{\SO}{{\rm SO}}

\newcommand{\blanc}{\ \ \ \ \ \ \ }

\newcommand{\N}{{\cal N}}
\newcommand{\CS}{{\rm CS}}

\begin{document}
\preprint{ITEP-TH-81/09}

\title{SQM WITH NON-ABELIAN SELF-DUAL FIELDS: HARMONIC SUPERSPACE DESCRIPTION}

\author{{\bf  Evgeny A. Ivanov},}

\affiliation{Bogoliubov Laboratory of Theoretical Physics, JINR, 141980
Dubna, Russia}

\author{{\bf Maxim A. Konyushikhin}, {\bf Andrei V. Smilga}}

\affiliation{SUBATECH, Universit\'e de Nantes, 4 rue Alfred Kastler, BP 20722, Nantes 44307, France}

\affiliation{
Institute for Theoretical and Experimental Physics,B. Cheremushkinskaya 25, Moscow 117259, Russia}

\begin{abstract}
\vskip 1cm\
\noindent We present a Lagrangian formulation
for  \mbox{${\cal N}=4$} supersymmetric
quantum-mechanical systems describing the motion in external
non-Abelian self-dual gauge fields. For any such system, one can write a component supersymmetric Lagrangian
by introducing extra bosonic variables with topological Chern-Simons type interaction. For a special class 
of such system when 
the fields are expressed in the  `t Hooft ansatz form, it is possible to give a superfield description using  
harmonic superspace formalism. As a new explicit example, the
${\cal N}=4$ mechanics with Yang monopole is constructed.
\end{abstract}


\maketitle

 \section{Introduction}

Supersymmetric quantum mechanics (SQM) provides a proper venue for exploring and modelling salient features
of supersymmetric field theories in diverse dimensions \cite{W}. Some SQM models are, in turn, $d=1$ reduction
of higher-dimensional supersymmetric theories. At the same time, many interesting models of this kind
can be constructed directly in \mbox{(0+1)} dimensions, without any reference to the dimensional reduction procedure.
They exhibit some surprising properties related to the peculiarities of $d=1$ supersymmetry. For any SQM model (like for any
supersymmetric field theory), it is desirable, besides the component formulation, to find out
the appropriate superfield Lagrangians. They make supersymmetry manifest,  prompt
possible generalizations of the model, and allow one to reveal relationships with other cognate theories.
The basic aim of the present paper is to construct such a Lagrangian formulation for a wide class of ${\cal N}=4$
SQM models,\footnote{Hereafter, $\N$ counts the number of {\sl real} supercharges.} with self-dual non-Abelian gauge field backgrounds.
The natural and necessary device for this formulation proves to be the harmonic superspace (HSS)
approach \cite{HSS} adapted to the one-dimensional case in \cite{IvLecht}.

The SQM models considered in this paper represent a subclass of the wider well-known class of system
that describes the motion of
a fermion on an even-dimensional manifold with an arbitrary gauge
background.
It was observed many years ago that one can treat this system 
as a supersymmetric one \cite{Gaume}. 
The
corresponding supercharges and the Hamiltonian are
\begin{equation}
\label{QHchir}
 Q = /\!\!\!\!     \D (1 + \gamma_5) ,
\blanc
 \bar Q = /\!\!\!\!     \D (1 - \gamma_5)  ,
\blanc
 H =  /\!\!\!\!     \D^2  .
\end{equation}
 Indeed, for any eigenstate $\Psi$ of the massless Dirac operator  $\, /\!\!\!\!     \D $
with a nonzero eigenvalue $\lambda$,
the state $\gamma^5 \Psi$ is also an eigenstate of  $\, /\!\!\!\!     \D $  with the eigenvalue
$-\lambda$. Thus, all excited states of $H$ are doubly degenerate.

For a four-dimensional flat manifold and self-dual Abelian or non-Abelian gauge fields, the spectrum of $H$ is 4-fold
degenerate implying extended ${\cal N} = 4$ supersymmetry.
For a flat Dirac operator in the instanton background, this can be traced back to Ref.~\cite{Jackiw}.
In Ref.~\cite{IvLecht},
 a \mbox{${\cal N}=4$} supersymmetric generalization of this system
describing the motion on a conformally flat $4D$ manifold with an Abelian self-dual background was found
and its off-shell Lagrangian formulation in the $d=1$ harmonic superspace was presented. In Ref. \cite{KonSmi},
it was noticed that a similar generalization exists for non-Abelian fields. The corresponding supercharges and the Hamiltonian have the form
 \begin{equation}\label{eq_Qf}
\begin{array}{l}
  Q_\alpha = f \left(\sigma_\mu \bar\psi\right)_\alpha \left(\hat p_\mu-\ca A_\mu\right)
-\psi_{\dot\gamma} \bar\psi^{\dot\gamma} \left(\sigma_\mu\bar\psi\right)_\alpha i\partial_\mu f,
\\[2mm]
  \bar Q^\alpha = \left(\psi\sigma^\dagger_\mu\right)^\alpha \left(\hat p_\mu-\ca A_\mu\right)f
+i\partial_\mu f \left(\psi\sigma^\dagger_\mu\right)^\alpha \psi_{\dot\gamma}\bar\psi^{\dot\gamma},
\end{array}
\end{equation}
\begin{multline}\label{eq_susyham}
  H=\frac{1}{2}f \left(\hat p_\mu-\ca A_\mu\right)^2 f
      +\frac{i}{4}f^2\, {\cal F}_{\mu\nu}\,\psi\sigma^\dagger_{\mu}\sigma_{\nu}\bar\psi
\\[2mm]
      - \frac 12 f i\partial_\mu f \,(\hat p_\nu-\ca A_\nu)\, \psi
    \sigma^\dagger_{[\mu} \sigma_{\nu]}\bar\psi
    + f\partial^2 f\left\{\psi_{\dot\gamma}\bar\psi^{\dot\gamma}-\frac{1}{2}\left(\psi_{\dot\gamma} \bar\psi^{\dot\gamma}\right)^2\right\}
\end{multline}
with self-dual gauge field ${\cal A}_\mu$ (Abelian or non-Abelian),
$\F_{\mu\nu}=\partial_\mu\A_\nu-\partial_\nu\A_\mu-i\left[\A_\mu,\A_\nu\right]$.
Complex fermion variables $\psi_{\dot{\alpha}}$ have two components,\footnote{We use the following SO(4) spinor notation:
$$
(\sigma_\mu)_{\alpha \dot\beta} = \left\{i, \sigma_a \right\}_{\alpha\dot\beta }, \;
\left(\sigma_\mu^\dagger\right)^{\dot\beta\alpha}=\left\{-i,\sigma_a\right\}^{\dot\beta\alpha} =
-\varepsilon^{\dot\beta\dot\gamma}\varepsilon^{\alpha\gamma}(\sigma_\mu)_{\gamma\dot\gamma}\,,
\quad \varepsilon_{\alpha\beta} = -\varepsilon^{\alpha\beta}, \varepsilon_{\dot\alpha\dot\beta} =
-\varepsilon^{\dot\alpha\dot\beta}, \varepsilon_{12}=1\,,
$$
$$
v_{\alpha \dot \alpha} =  v_\mu (\sigma_\mu)_{\alpha \dot\alpha}\ ,\ \ \ \
v_\mu = \frac 12 v_{\alpha \dot\alpha}
(\sigma_\mu^\dagger)^{\dot \alpha \alpha}\, =\, -\frac 12 v^{\alpha \dot\alpha}
(\sigma_\mu)_{\alpha \dot\alpha}\,, \;
v^{\alpha \dot \alpha} = \varepsilon^{\alpha \beta}\varepsilon^{\dot\alpha\dot\beta } v_{\beta \dot\beta}
\, =\,  -v_\mu (\sigma_\mu^\dagger)^{\dot\alpha \alpha}\,.$$}
$\hat p_\mu = -i\partial/\partial x_\mu$ and $f(x)$ is an arbitrary scalar function determining the conformally flat metric,
$ds^2 = \left\{f(x)\right\}^{-2} dx_\mu^2$. Note that the supercharges and the Hamiltonian written above are $\SO(4)=\SU(2)\times\SU(2)$ covariant;
the undotted indices come from the first $\SU(2)$ (R-symmetry group), while the dotted ones -- from the second $\SU(2)$
commuting with supersymmetry.

When the gauge field $\A_\mu$ is Abelian, the corresponding Lagrangian can be easily written,
\begin{multline}\label{eq_action}
     L \ =\
    \frac{1}{2}f^{-2}\, \dot x_\mu\dot x_\mu
     +i{\bar\psi}^{\dot\alpha} \dot{\psi}_{\dot\alpha}
    +\frac{1}{4} \left\{3\left(\partial_\mu f\right)^2- f\partial^2 f \right\} \psi^4
     +\frac i2 f^{-1}\partial_\mu f \,\dot x_\nu\,
      \psi  \sigma^\dagger_{[\mu}\sigma_{\nu]} \bar\psi
\\
    +\ca A_\mu(x)\dot x^\mu
    - \frac{i}{4}f^2 {\cal F}_{\mu\nu}\,\psi\sigma^\dagger_{\mu}\sigma_{\nu}\bar\psi,
\end{multline}
where the second line represents the interaction term with the Abelian gauge field.
This Lagrangian admits a superfield formulation  \cite{IvLecht,KonSmi} in the framework of the
harmonic superspace (HSS) approach \cite{HSS}.

For a matrix-valued non-Abelian self-dual field  ${\cal A}_\mu$, the (scalar) Lagrangian cannot be straightforwardly derived from
(\ref{eq_susyham}). We will show
that this can be done by introducing extra ``semi-dynamical'' fields $\varphi_i$
in the fundamental representation of $\SU(N)$ and the auxiliary ${\rm U}(1)$ gauge field $B(t)$.
The second line in (\ref{eq_action}) is then generalized to
\begin{equation}\label{eq_lagrNA}
 L_{\rm int}^{\SU(N)}=i\bar\varphi^i\left(\dot\varphi_i+iB\varphi_i\right)
  +kB
  +\A_\mu^a T^a \,\dot x_\mu-\frac{i}{4}f^2\F_{\mu\nu}^a T^a\,\psi\sigma_\mu^\dagger\sigma_\nu\bar\psi
\end{equation}
with $k$ integer and
\begin{equation}\label{eq31}
T^a=\bar\varphi^i\left(t^a\right)^{\,\, j}_{\! i} \varphi_j,
\end{equation}
$t^a$ being standard $\SU(N)$ algebra generators.
The interaction Lagrangian (\ref{eq_lagrNA}) is $\N=4$ supersymmetric.
The corresponding supersymmetry transformations are written in Eqs.~(\ref{susytran}).
It is easy to check that it is covariant with respect to the target space non-Abelian gauge transformations
\begin{equation}\label{gauge}
\begin{array}{c}
 \A_\mu^a\, t^a\rightarrow U^\dagger \A_\mu^a\, t^a U+iU^\dagger\partial_\mu U
\\[2mm]
 \varphi_i\rightarrow\left(U^\dagger\varphi\right)_i,\,\,\,\,
 \bar\varphi^i\rightarrow \left(\bar\varphi U\right)^i
\end{array}
\end{equation}
with $U(x)\in\SU(N)$.

It is not immediately clear how to extend the Abelian superfield
description to  the general non-Abelian case, i.e. to the gauge group
${\rm SU}(N)\,$.
In this paper, we construct such a description for the particular case
of $\SU(2)$ self-dual fields expressed in the form
 \begin{equation}
\label{Hooft} {\rm (a)}\;\;{\cal A}^a_\mu \ =\ -\bar \eta^a_{\mu\nu}
\partial_\nu \ln h(x) \ \ \ \ \ \ {\rm or}\ \ \ \ \ \ \ {\rm (b)}
\;\;{\cal A}^a_\mu \ =\ -\eta^a_{\mu\nu} \partial_\nu \ln h(x)\, ,
 \end{equation}
with harmonic function $h(x)$,
$$\partial_\mu^2 h(x) = {\rm a\ finite\ sum\ of \ delta\ functions}$$
(the expressions (\ref{Hooft}a) and (\ref{Hooft}b) correspond, respectively, to self-dual and anti-self-dual fields).
This is the so called 't Hooft ansatz for a multi-instanton $\SU(2)$ solution \cite{Hooft}, the symbols $\eta^a_{\mu\nu}$
being defined as
  \begin{equation}\label{tHooft}
\eta^a_{ij} = \bar\eta^a_{ij} = \varepsilon_{aij},\ \ \eta^a_{i0} =
- \eta^a_{0i} = \bar\eta^a_{0i} =  -\bar\eta^a_{i0} =  \delta_{ai}
  \blanc\left(i,j,a=1,2,3\right) .
\end{equation}

For generic self-dual ADHM \cite{ADHM} configurations, the problem of finding a superfield Lagrangian is more complicated.
This problem is under study now.

\section{Derivation}
 In the ${\cal N}=4$, $d=1$ HSS approach \cite{IvLecht}, the superfields depend on bosonic variables $t,\  u^{\pm \alpha}$ (the harmonics
$u^{+\alpha}, u^{-}_\alpha = (u^{+\alpha})^*$ satisfying the constraint
$u^{+\alpha} u^-_\alpha = 1$
parametrize the R-symmetry group $\SU(2)$ of the ${\cal N} = 4$
superalgebra)  and on fermionic variables $\theta^\pm = u_\alpha
^\pm \theta^\alpha $, $\bar\theta^\pm = u_\alpha ^\pm
\bar\theta^\alpha $. The most striking feature of HSS is the
presence of an {\sl analytic superspace} $\left\{t_{\rm A},
\theta^+, \bar\theta^+, u^{\pm \alpha}\right\}$ in it (an analog of
${\cal N} = 2$ chiral superspace) involving the  ``analytic time''
$t_A = t + i(\theta^+ \bar \theta^- + \theta^- \bar \theta^+)$ and
containing twice as less fermionic coordinates. Our definitions and
conventions are the same as in \cite{KonSmi} (and similar to
\cite{IvLecht}; see \cite{HSS} for more details) and we mostly will
not repeat them here.

To construct the action, introduce, following \cite{IvLecht,KonSmi},  a
 doublet of superfields $q^{+{\dot\alpha}}$ with charge +1 ($D^0 q^+ = q^+$)
satisfying the constraints
\begin{equation}\label{eq_anal}
  D^+ q^{+{\dot\alpha}} =0,\blanc
  \bar D^+ q^{+{\dot\alpha}} = 0,\blanc
  D^{++}q^{+{\dot\alpha}} =0 \,,
\end{equation}
where $D^+, \ \bar D^+$ and $D^{++}$ are spinor and harmonic
derivatives.\footnote{The constraints $D^+ q^+ = \bar D^+ q^+ = 0$ are akin to widely known chirality
constraints
like $D_\alpha \Phi = 0$ in ${\cal N} =1$, $d=4$ supersymmetric theories. In the analytic basis,
they simply mean that $q^+$ does not depend on $\theta^-, \bar \theta^-$. Such constraints appear
naturally in  the HSS formalism and are common also in $d=4$ theories. A possibility to impose
the extra  constraint $D^{++} q^+ = 0$ is specific for the (0+1)-dimensional case,
where it has a pure kinematic nature. In $\N=2$, $d=4$ theories, the relation $D^{++} q^+ = 0$ is not
a kinematic constraint, it is the equation of motion for the {\it free} hypermultiplet
derived from the action $S = \int d^4x \, du \, d^4\theta^+ \, \widetilde {q^+} D^{++} q^+$ \cite{HSS}.}
We impose also the pseudoreality condition
     \begin{equation}
  \label{reality}
   \widetilde{q^{+{\dot\alpha}}} =   q^{+}_{\ \dot\alpha} =\varepsilon_{{\dot\alpha}{\dot\beta}} q^{+{\dot\beta}}\ ,
   \blanc \widetilde{\widetilde{q^{+{\dot\beta}}}} =
    - q^{+{\dot\beta}}\,,
    \end{equation}
where the field $\widetilde{q^+}$ is obtained from $q^+$ by an involution transformation.
{\it On top of that}, we introduce an analytic gauge superfield $V^{++}$  of charge +2 satisfying the constraints
\begin{equation}
\label{constrV}
 D^+ V^{++}  = \bar D^+ V^{++} = 0 \ ,\blanc  V^{++} = \widetilde {V^{++}}
\end{equation}
and the ``matter'' superfield $v^+$ of charge +1.  The  constraints it satisfies,
  \begin{equation}\label{covarconstr}
  D^+ v^{+} =0,\blanc
  \bar D^+ v^{+} = 0,\blanc
  (D^{++} +i V^{++} ) v^{ +} =0 \, ,
\end{equation}
 differ from (\ref{eq_anal}) by the presence of the covariant harmonic derivative ${\cal D}^{++} = D^{++} + iV^{++}$
\cite{Delduc}.
The constraint ${\cal D}^{++} v^+ = 0$ is covariant with respect to gauge transformations
\begin{equation}
\label{gaugeharm}
V^{++} \ \to \ V^{++} + D^{++} \Lambda,\blanc
 v^+ \ \to\ e^{-i\Lambda} v^+ ,
\blanc
D^+\Lambda=\bar D^+\Lambda=0\ .
\end{equation}

The constraints (\ref{eq_anal}), (\ref{covarconstr}) drastically reduce the number of the physical component fields in the superfields
$q^{+{\dot \alpha}}$ and $v^+$. The solution of (\ref{eq_anal}), (\ref{reality}) for $q^{+{\dot \alpha}}$ in the analytic basis, with
\begin{equation}
 D^{++} = \partial^{++} + 2i\theta^+\bar\theta^+\frac{\partial}{\partial t_{\rm A}}\,,
  \blanc
 \partial^{++} = u^{+}_\alpha\frac{\partial}{\partial u^{-}_\alpha}\,,
\end{equation}
is
\begin{equation}\label{eq_qdot}
  q^{+\dot\alpha} = x^{\alpha\dot\alpha}(t_{\rm A})u^+_\alpha
  -2\theta^+\chi^{\dot\alpha}(t_{\rm A})-2\bar\theta^+\bar\chi^{\dot\alpha}(t_{\rm A})
  -2i\theta^+\bar\theta^+  \dot x^{\alpha \dot\alpha}  u^-_\alpha \ ,
\end{equation}
 where
\begin{equation}
  x^{\alpha{\dot\alpha}}=-\left(x_{\alpha{\dot\alpha}}\right)^* ,\blanc
  \bar\chi^{\dot\alpha}=\left(\chi_{\dot\alpha}\right)^*\ .
\end{equation}
The first condition in the latter equation means that
$x_\mu = -\frac 12x^{\alpha \dot\alpha} (\sigma_\mu)_{\alpha \dot\alpha}$ is real, and we are left with four dynamic bosonic variables.

We can use the gauge freedom (\ref{gaugeharm}) to eliminate almost all components from $V^{++}$ and to present it as
\begin{equation}
\label{VPP}
V^{++} \ =\ 2i\, \theta^+ \bar \theta^+ B,
\end{equation}
where the gauge field $B(t)$ is real.
This is a $d=1$ counterpart of the familiar Wess-Zumino gauge in $d=4$ theories.
Then the superfield $v^+$ is expressed in the analytical basis as
\begin{equation}
\label{vp}
  v^+=\phi^\alpha u^+_\alpha
  -2\theta^+\omega_1-2\bar\theta^+\bar\omega_2
  -2i\theta^+\bar\theta^+  (\dot \phi^\alpha + i B\phi^\alpha) u^-_\alpha\ ,
\end{equation}
from which it follows that
\begin{equation}
\label{vptild}
 \widetilde  {v^+}= \bar \phi^\alpha u^+_\alpha
  -2\theta^+\omega_2  + 2\bar\theta^+ \bar\omega_1
  -2i\theta^+\bar\theta^+ ({\dot {\bar \phi}^\alpha} - iB \bar \phi^\alpha) u^-_\alpha\, \
\end{equation}
with $\bar \phi^\alpha = (\phi_\alpha)^*$.
Thus, the fields $\phi_\alpha$ and $\bar\phi^\alpha$ are charged under ${\rm U}(1)$ gauge field $B$ and have opposite charges.

The  $\ca N=4$ SUSY invariant action that we are going to write consists of three parts, $S=S_{\rm kin}+S_{\rm int} + S_{\rm FI}$.
The kinetic part is more convenient to express in the central basis
$\{t, \theta_\alpha, \bar\theta^\beta\}$.
It has  the same form as in \cite{IvLecht,KonSmi},
\begin{equation}
\label{Lkin}
  S_{\rm kin}=\int  dt\, d^4\theta\, du\, R'_{\rm kin}(q^{+{\dot\alpha}}, q^{-{\dot\beta}},
u^\pm_\gamma)=\int dt\,d^4\theta\, R_{\rm kin}(q_\mu),
\end{equation}
where $q_\mu=-\frac 12 q^{\alpha\dot\alpha}\left(\sigma_\mu\right)_{\alpha\dot\alpha}$
(the equation $D^{++}q^{+\dot\alpha}=0$ ensures that in the central basis $q^{+\dot\alpha}$ depends on $u^\pm_\beta$ linearly,
i.e. $q^{+\dot\alpha}=q^{\alpha\dot\alpha}u^+_\alpha$; see also Eq.~(35) in \cite{KonSmi}),
with an arbitrary function $R_{\rm kin}(q_\mu)$ of the real superfield $q_\mu$.
The component expansion of (\ref{Lkin}) coincides with the first line in Eq.~(\ref{eq_action}), where
$f(x)=\left[\frac 12\partial_\mu^2 R_{\rm kin}\right]^{-1/2}$ and \mbox{$\psi_{\dot\alpha}=f^{-1}\chi_{\dot\alpha}$} \cite{KonSmi}.

The interaction part is chosen as
\begin{equation}
\label{Sint}
 S_{\rm int}= -\frac 12 \int \, dt \, du\, d\bar \theta^+d\theta^+\,
  K\left(q^{+\dot{\alpha}}, u^\pm_\beta\right)  v^+ \widetilde{v^+} \ ,
\end{equation}
where the condition $\widetilde K=K$ is imposed to ensure the action to be real.
Finally, we add the Fayet-Illiopoulos term
\begin{equation}
\label{FI}
S_{\rm FI} \ =\ -\frac {i k}2  \int \, dt \, du \, d\bar \theta^+ d\theta^+ \, V^{++}\
=\ k\int dt\,  B \  ,
\end{equation}
which is invariant under gauge transformations (\ref{gaugeharm}). At the classical level, $k$ is an arbitrary real number. As we will
shortly see, a benign quantum theory can only be defined if the requirement
 \begin{equation}
\label{kvantovanie}
k \ =\ {\rm integer}
 \end{equation}
is fulfilled.

Let us concentrate on the interaction part. It is convenient to introduce new variables
 \begin{equation}
\label{varphi}
\varphi_\alpha \ =\ \phi_\alpha \sqrt {h(x) }\ ,
 \end{equation}
 where
\begin{equation}\label{h}
h(x) = \int du\, K(x^{+\dot{\alpha}}, u^\pm_\beta ),
\blanc
 x^{+\dot\alpha}=x^{\alpha\dot\alpha}u_\alpha^+,
\end{equation}
is a harmonic function.\footnote{We assumed here that $h(x)>0$.}   Indeed,
$$\partial_\mu^2 h(x) = 4\varepsilon^{\dot\alpha\dot\beta}\int du\,
\partial_{+\dot\alpha}\partial_{-\dot\beta}K(x^{+\dot\gamma}, u^\pm_\beta) = 0\,.
$$

Substituting (\ref{eq_qdot}), (\ref{vp}) and (\ref{vptild}) into (\ref{Sint}) and eliminating the auxiliary  fermionic degrees
of freedom $\omega_{1,2}$, $\bar \omega_{1,2}$ by their algebraic equations of motion, we derive after some algebra
\begin{equation}
\label{Lint}
 L_{\rm int} \ =\
 i \bar \varphi^\alpha ({\dot \varphi}_\alpha + iB \varphi_\alpha)
 -  \frac 12 \bar \varphi^\beta \varphi_\gamma
\left({\cal A}_{\alpha{\dot\alpha}} \right)_{\!\beta}^{\,\,\gamma} {\dot x}^{\alpha {\dot \alpha}}
  - \frac i4 \left( {\cal F}_{{\dot\alpha}{\dot\beta}}\right)^{\,\,\gamma}_{\!\beta}
 \chi^{\dot\alpha}
\bar \chi^{\dot\beta} \bar\varphi^\beta \varphi_\gamma
\,. \end{equation}
Here
\begin{equation}
\label{Adef}
\left({\cal A}_{\alpha{\dot\alpha}} \right)_{\!\beta}^{\,\,\gamma}  = -\frac {2i}{\int du \, K} \int du \, \partial_{+\dot\alpha}
K \left( u^{+\gamma}\varepsilon_{\alpha\beta}
  - \frac 12
u^+_\alpha \delta^\gamma_\beta \right)
=\frac{i}{h}
\left(\varepsilon_{\alpha\beta} \, \partial^\gamma_{\ \dot\alpha} h - \frac 12
 \delta^\gamma_\beta \,\partial_{\alpha{\dot\alpha}}h  \right)
\end{equation}
($\partial_{\alpha\dot\alpha}\equiv \left(\sigma_\mu\right)_{\alpha\dot\alpha}\partial_\mu =-2\partial/\partial x^{\alpha\dot\alpha}$)
is a Hermitean traceless matrix, the gauge field, and
\begin{equation}
\label{Fdef}
\left( {\cal F}_{{\dot\alpha}{\dot\beta}} \right)^{\,\,\gamma}_{\!\beta}
= \left(\F_{\mu\nu}\right)^{\,\,\gamma}_{\!\beta}
    \left(\sigma_\mu^\dagger\sigma_\nu\right)_{\dot\alpha\dot\beta}
  =\partial_{\delta\dot\alpha}
( {\cal A}^{\delta}_{\ {\dot\beta}} )^{\,\,\gamma}_{\!\beta}
-i ( {\cal A}_{\delta {\dot\alpha} } )^{\,\,\lambda}_{\!\beta}
  ( {\cal A}^{\delta}_{\ {\dot\beta} } )^{\,\,\gamma}_{\!\lambda} \   +
({\dot\alpha} \leftrightarrow {\dot\beta} )
 \end{equation}
is its self-dual part. It is easy to check explicitly, that the anti-self-dual part of the gauge field $\A_\mu$ vanishes,
\begin{equation}\label{Fanti}
 \left(\F_{\alpha\beta}\right)_{\!\gamma}^{\,\,\delta}
  =\left(\F_{\mu\nu}\right)_{\!\gamma}^{\,\,\delta} \left(\sigma_\mu\sigma_\nu^\dagger\right)_{\alpha\beta}
    =-\partial_{\alpha\dot\alpha}(\A_\beta^{\ \dot\alpha})^{\,\,\delta}_{\!\gamma}
    +i(\A_{\alpha\dot\alpha})^{\,\,\lambda}_{\!\gamma}
      (\A_\beta^{\ \dot\alpha})^{\,\,\delta}_{\!\lambda}
        +\left(\alpha\leftrightarrow\beta\right)
    =0.
\end{equation}
Thus, the field strength ${\cal F}_{\mu\nu}^a$ is self-dual and belongs
to the representation $(0,1)$ of ${\rm SO}(4)\,$. Passing to ${\cal A}_\mu^a$
as $\left(\A_\mu\right)^{\,\,\gamma}_{\!\beta}=\A_\mu^a \left(\sigma_a\right)^{\,\,\gamma}_{\!\beta}\!\!/2\,$,
we find that the representation \p{Adef} precisely amounts to the self-dual 't~Hooft ansatz (\ref{Hooft}a).
The anti-self-dual expression (\ref{Hooft}b) arises if one interchanges altogether dotted and undotted indices,
i.e. effectively interchanges $\sigma_\mu$ and $\sigma^\dagger_\mu$. This also implies passing to the harmonics
$u^\pm_{\dot\alpha}$ and in fact to another ${\cal N}=4$ supersymmetry, with the second SU(2) (acting on dotted indices)
as the R-symmetry group.

Finally, substituting
$\bar\varphi^\beta\varphi_\gamma=T^a\left(\sigma_a\right)_{\!\gamma}^{\,\,\beta}$
and $\chi_{\dot\alpha}=f\psi_{\dot\alpha}$ into (\ref{Lint}), where $T^a$ is defined in (\ref{eq31}) with $t^a=\frac 12\sigma_a$,
we convince ourselves that the interaction term together with the FI term \p{FI} yields just (\ref{eq_lagrNA}) for the SU(2) case.
The canonical Hamiltonian derived from the Lagrangian $L_{\rm kin} + L_{\rm int} + L_{\rm FI}$ has the form (\ref{eq_susyham})
with ${\cal A}_\mu \equiv {\cal A}_\mu^a T^a$ and ${\cal F}_{\mu\nu} \equiv   {\cal F}_{\mu\nu}^a T^a $.

Observe that the variables $\varphi_\alpha$ enter the Lagrangian with only one time derivative. Thus, they are
 not full-fledged dynamic variables (like $x_\mu$) and not auxiliary fields (like $\omega_{1,2}$). They have a kind
 of intermediate nature.\footnote{In the context of ${\cal N}=4$ SQM models, such variables (together
 with their analytic superfield carriers $v^+, \widetilde{v^+}$) were introduced in \cite{fil1,fil2}
 (for a recent application, see also \cite{bks}).}
To understand it better, perform the quantization. To begin with, it is sufficient to restrict oneself by the first term
in (\ref{Lint}) with addition of the Fayet-Illiopoulos term (\ref{FI}).
The action
 \begin{equation}
 \label{SphiB}
 S \ =\ \int dt\, \Big[ i \bar \varphi^\alpha (\dot{\varphi}_\alpha + iB \varphi_\alpha) \ + \ kB \Big]
 \end{equation}
much resembles the $3D$ Chern-Simons action,
 \begin{equation}
\label{CS}
S_{\CS} = \kappa \int \left( A \wedge dA - \frac {2i}3 A\wedge A \wedge A \right)\ .
 \end{equation}
In both systems, the canonical Hamiltonian is zero, the canonical momenta are algebraically expressed through coordinates,
 and the quantization consists in imposing certain second class constraints (for a nice review of the classical and quantum aspects of
the Chern-Simons theory, see \cite{Dunne}).
Another well-known feature of CS theory is the quantization of the coupling, $k_{\CS} = 4\pi \kappa$ = integer. This follows from the requirement
for the Euclidean path integral to be invariant with respect to large (topologically nontrivial) gauge transformations. As was mentioned above, in
our case the coefficient $k$ is also quantized. This can be derived following a similar reasoning.

Notice first that the action (\ref{SphiB}) is invariant with respect to gauge transformations,
 \begin{equation}
\label{gaugeM}
B(t) \to \ B(t) + \frac {d \alpha(t)}{dt}, \ \ \ \ \ \ \ \ \ \ \varphi(t) \to e^{-i\alpha(t)} \varphi (t) \ ,
 \end{equation}
which, in the Euclidean version of the theory, become
 \begin{equation}
\label{gaugeE}
B(\tau) \to \ B(\tau) + i \frac {d\alpha(\tau)}{d\tau} , \ \ \ \ \ \ \ \ \ \ \varphi(\tau) \to e^{-i\alpha(\tau)} \varphi (\tau) \ .
 \end{equation}
This is a remnant of  gauge transformations (\ref{gaugeharm}), which survives
in the Wess-Zumino gauge (\ref{VPP}).
To discover topologically nontrivial gauge transformation, consider the Euclidean version
of this theory and regularize it in the infrared by putting it on a finite Euclidean interval
$\tau \in (0,\beta)$ and imposing the periodic boundary conditions
$B(\beta) = B(0), \ \varphi(\beta) = \varphi(0)$.\footnote{This is of course equivalent to introducing a finite temperature $T = 1/\beta$.}
  Then the only admissible gauge transformations (\ref{gaugeE}) are those which do not break these periodicity conditions.
We see that the  transformation  with $\alpha(\tau) = 2\pi \tau/\beta$ is topologically nontrivial, it cannot be reduced to a chain
of infinitesimal transformations. This transformation shifts the Euclidean Fayet-Illiopoulos action by an {\it imaginary} constant,
$\Delta S_{\rm FI} = -2\pi i k$. The requirement that the Euclidean path integrals (involving the factor $e^{-S_{\rm FI}}$) are not changed
leads \cite{Polych} to the quantization condition (\ref{kvantovanie}).

The fact that $k$ must be integer leads to the finite representations of the operator algebra $\varphi_\alpha$, $\bar\varphi^\alpha$.
Indeed, consider the integer $k$ to be positive as required in the classical case due to the constraint
\begin{equation}
\bar\varphi^\alpha\varphi_\alpha = k\,, \lb{classConstr}
\end{equation}
which follows from \p{SphiB} by varying with respect to $B$ (at the quantum level, negative $k$ can be equally chosen).
The canonical commutation relations following from the same action (\ref{SphiB}) through the standard Dirac prescription  are
\begin{equation}
\label{phicomm}
 [ \varphi_\alpha, \bar \varphi^\beta] = \delta_\alpha^\beta \ , \ \ \ \ \ \ \ \ \ \ \ \
[\varphi_\alpha, \varphi_\beta] = [\bar \varphi^\alpha, \bar \varphi^\beta] = 0 \ .
 \end{equation}
In quantum theory, one can choose $\varphi_\alpha \equiv \partial/\partial \bar\varphi^\alpha$ and impose \p{classConstr} on the wave functions:
 \begin{equation}
 \label{svjazk}
 \bar \varphi^\alpha \varphi_\alpha \Psi =  \bar\varphi^\alpha \frac \partial {\partial \bar\varphi^\alpha} \Psi  = k\Psi\ .
 \end{equation}
In other words, the wave functions represent homogeneous polynomials of $\bar\varphi^\alpha$ of (an integer) degree $k\,$.\footnote{In the case $k<0$
the algebra (\ref{phicomm}) is the same. One must
choose $\bar\varphi^\alpha= - \partial/\partial\varphi_\alpha$ and consider polynomials of $\varphi_\alpha$ of degree $|k|$.}
The number of such (linearly independent) polynomials is $k+1$.
 Moreover, it is also easy to see that the operators (\ref{eq31}) satisfy the following algebra
 \begin{equation}
\label{algTa}
[T^a, T^b] \ =\ i\varepsilon^{abc} T^c\ .
 \end{equation}
In addition, taking into account (\ref{svjazk}), we derive
 \begin{equation}
\label{Kasimir}
T^a T^a \ =\ \frac 14 \left[ (\bar\varphi^\alpha \varphi_\alpha)^2 + 2 (\bar\varphi^\alpha \varphi_\alpha) \right] \
=\  \frac k2 \left(\frac k2 +1\right) .
 \end{equation}
In other words, $T^a$ can be treated as the generators of $\SU(2)$ in the representation
of spin $k/2$.\footnote{This way of quantizing semi-dynamical variables $\varphi_\alpha, \bar\varphi^\alpha$
was employed in Ref.~\cite{fil2}. Alternatively, one could interpret
$\varphi_\alpha, \bar\varphi^\alpha$ with the constraint \p{classConstr} as a kind of the target harmonic
variables representing a sphere ${\rm S}^2$, solve \p{classConstr} in terms of the stereographic projection coordinates
$z(t)$ and $\bar z(t)$, and quantize the system by the Gupta-Bleuler method as in Ref.~\cite{cpnm}.}
The nice feature is that this gauge $\SU(2)$ is in fact R-symmetry group of $\N=4$ supersymmetry algebra.

\section{Discussion and outlook}
Our main result is the HSS superfield 
action for the ${\cal N}=4$ SQM
corresponding to the Hamiltonian (\ref{eq_susyham}) with a non-Abelian $\SU(2)$ gauge field ${\cal A}_\mu$ which
lives on a conformally flat 4-manifold and is representable in the 't Hooft ansatz form (\ref{Hooft}).

As an example of such a field, we can quote the instanton solution on ${\rm S}^4$. Generically, it depends on the radius $R$ of the sphere
and the instanton size $\rho$. The configurations of maximal size, $\rho = R$, present a particular interest.
In the stereographic coordinates on ${\rm S}^4$,
 \begin{equation}
\label{metricS4}
 ds^2 \ =\ \frac {4R^4 dx_\mu^2}{(x^2 + R^2)^2}\ ,
 \end{equation}
they are expressed by the same formulae as flat instantons in singular gauge,
  \begin{equation}
   \label{inst}
  {\cal A}_\mu^a \ =\ \frac{2R^2 \bar\eta^a_{\mu\nu} x_\nu }{x^2(x^2+ R^2)} \quad \mbox{or} \quad
  ({\cal A}_{\alpha\dot\alpha})^{\,\,\gamma}_{\!\beta} = - \frac{2i \,R^2}{x^2(x^2 + R^2)}\left(\varepsilon_{\alpha\beta}x^\gamma_{\dot\alpha}
  - \frac{1}{2}\,\delta^\gamma_\beta \,x_{\alpha\dot\alpha} \right),
 \end{equation}
and
\begin{equation}
({\cal F}_{\dot\alpha\dot\beta})^{\,\,\gamma}_{\!\beta} = \frac{8i \,R^2}{x^2(x^2 + R^2)^2}\left(x^\gamma_{\dot\beta}x_{\beta\dot\alpha}
+ x^\gamma_{\dot\alpha}x_{\beta\dot\beta}\right). \lb{FstSing}
\end{equation}
The corresponding functions in Eq.~(\ref{h}) are taken in the form
\begin{equation}\label{Kh}
 K(x^{+\dot\alpha},u^\pm_\beta)
  =1+\frac{1}{\left(c^-_{\dot\alpha}x^{+\dot\alpha}\right)^2}\ ,
\blanc
 h(x)\equiv \int du\, K(x^{+\dot\alpha},u^\pm_\beta)=1+\frac{R^2}{ x_\mu^2},
\end{equation}
where $c^-_{\dot\alpha}=c_{\ \dot\alpha}^\alpha u^-_\alpha$, $c^{\alpha\dot\alpha}$ --
constant vector and $R^2=1/c^2_\mu$.\footnote{The integral on the right hand side of Eq.~(\ref{Kh})
can be calculated as the power series in
$c^-_{\dot\alpha}c^{+\dot\alpha}=-c_\mu^2$ or directly after noting that this integral is ${\rm SO}(4)$
covariant and putting $c_\mu=(c,0,0,0)$, $x_\mu=(x_1,x_2,0,0)$.}
The field $\A_\mu^a$ can be brought to nonsingular gauge
\begin{equation}\label{non-singular}
 \A_\mu^a=\frac{2\eta_{\mu\nu}^a x_\nu }{x^2+R^2}\,, \qquad {\cal F}^a_{\mu\nu}
\ =\ - \frac {4R^2\eta^a_{\mu\nu}}{(x^2 + R^2)^2}\,,
\end{equation}
by the gauge transformations (\ref{gauge}) with $U(x)=-i\sigma_\mu
x_\mu/\sqrt{x^2}$ (this $U(x)$ is prompted by the form of the field
strength \p{FstSing}). The action density $\sim \F_{\mu\nu}
\F^{\mu\nu}$ is the same in this case at all points of ${\rm S}^4$. It is
worth noting that the singular gauge transformation converts the
undotted gauge group indices into the dotted ones: the self-dual
gauge potential and the field strength in the spinorial notation
become
\begin{equation}
({\cal A}_{\alpha\dot\alpha})^{\!\dot\gamma}_{\,\,\dot\beta} = \frac{2i}{x^2 + R^2}\left(\varepsilon_{\dot\alpha\dot\beta}x^{\dot\gamma}_{\alpha}
  - \frac{1}{2}\,\delta^{\dot\gamma}_{\dot\beta} \,x_{\alpha\dot\alpha} \right),\quad
({\cal F}_{\dot\alpha\dot\beta})^{\!\dot\gamma}_{\,\,\dot\delta} = -\frac{8i \,R^2}{(x^2 + R^2)^2}
\left(\varepsilon_{\dot\alpha\dot\delta}\delta^{\dot\gamma}_{\dot\beta} + \varepsilon_{\dot\beta\dot\delta}\delta^{\dot\gamma}_{\dot\alpha}\right)
\end{equation}
and, also, $\varphi_\alpha \rightarrow \varphi^{\dot\alpha} = -i\varphi_\alpha\,
x^{\alpha\dot\alpha}/\sqrt{x^2}$,
$\bar\varphi^\alpha \rightarrow \bar\varphi_{\dot\alpha} = -i\bar\varphi^\alpha\,
x_{\alpha\dot\alpha}/\sqrt{x^2} \,$.

Actually, the field (\ref{inst}), (\ref{non-singular}) describes Yang monopole living
in ${\mathbb R}^5$ \cite{Yang}. The potential \p{non-singular} has a nice group-theoretical meaning as one of the two SU(2)
connections on the coset manifold $\SO(5)/[\SU(2)\times \SU(2)] \sim {\rm S}^4$ (see e.g. \cite{gp}). It coincides with the flat self-dual instanton
only in the conformally flat parametrization of ${\rm S}^4$ as in \p{metricS4}. When coupled to the world-line through
our semi-dynamical variables $\varphi_\alpha, \bar\varphi^\alpha$, the 5-dimensional Yang monopole is reduced to this SU(2)
connection defined on ${\rm S}^4$.

Let us elaborate on this point in more detail, choosing, without loss of generality, $R =1$ in the above formulas.
Consider the following  $d=1$ bosonic Lagrangian with
the ${\mathbb R}^5$ target space and an additional coupling to Yang monopole
\begin{equation}
L_{\,{\mathbb R}^5} = \frac 12\left(\dot{y}_5 \dot{y}_5 +  \dot{y}_\mu \dot{y}_\mu\right) + \B_\mu^{\, a} (y) T^a \,\dot y_\mu \,.\lb{Yang1}
\end{equation}
Here, $\B_\mu^{\,a} $ is the standard form of the Yang monopole in the ${\mathbb R}^5$ coordinates,
\begin{equation}
\B_\mu^{\,a} = \frac{\eta_{\mu\nu}^a y_\nu }{r(r + y_5)}\,, \quad r = \sqrt{y_5^2 + y^2_\mu}\,,
\end{equation}
$T^a$ are defined as in \p{eq31} with $t^a = {\textstyle \frac{1}{2}}\sigma^a\,$, and we omitted the action for the semi-dynamical
variables $\varphi_\alpha, \bar\varphi^{\alpha}$. Now we pass to the polar decomposition of ${\mathbb R}^5$ into a radius $r$ and the angular
part ${\rm S}^4\,$, $(y_5, y_\mu) \;\rightarrow \; (r, \tilde{y}_5, \tilde{y}_\mu)\,, \; \tilde{y}_5 = \sqrt{1 - \tilde{y}^2_\mu}\,$,
and rewrite \p{Yang1} as
\begin{equation}
L_{\,{\mathbb R}^5} =
\frac 12\dot{r}{}^2 + \frac 12 r^2\left(\dot{\tilde{y}}_5 \dot{\tilde{y}}_5 + \dot{\tilde{y}}_\mu \dot{\tilde{y}}_\mu\right)
+ \frac{\eta_{\mu\nu}^a \tilde{y}_\nu \dot{\tilde{y}}_\mu\,T^a }{1 + \sqrt{1 - \tilde{y}^2_\mu}} \,.\lb{Yang2}
\end{equation}
The coordinates $\tilde{y}_\mu$ give a particular parametrization of ${\rm S}^4$. Passing to the stereographic coordinates is accomplished by
the redefinition
$$
\tilde{y}_\mu = 2\, \frac{x_\mu}{ 1 + x^2}\,,
$$
which casts \p{Yang2} into the form
\begin{equation}
L_{\,{\mathbb R}^5} =
\frac 12\left\{\dot{r}{}^2 + 4 r^2\frac{\dot{x}_\mu\dot{x}_\mu}{(1 + x^2)^2}
\right\}
+ \frac{2 \eta_{\mu\nu}^a x_\nu \dot{x}_\mu\,T^a }{1 + x^2} \,.\lb{Yang3}
\end{equation}
We see that the ${\rm S}^4$ metric \p{metricS4} (with $R =1$) and the instanton
vector potential \p{non-singular} appear.

Thus, our approach, as a by-product, provides a solution
to the long-standing problem of constructing ${\cal N}=4$ SQM with Yang monopole (see e.g. \cite{gknty} and references therein).
Obviously, the component Lagrangian (\ref{eq_action}) (with the relevant function $f(x)$) is just
the ${\rm S}^4$ part of the Lagrangian \p{Yang3} with the ``frozen'' radial variable $r = 1$. Presumably, one can restore the
full 5-dimensional kinetic part in \p{Yang3} by adding a coupling to the appropriately constrained scalar ${\cal N}=4$
zero-charge superfield $X(t,\theta, \bar\theta)$ which describes an off-shell multiplet $({\bf 1, 4, 3})$ with
one physical bosonic field \cite{IKLev}, such that $X|_{\theta = \bar\theta = 0} = r\,$.

The problem of finding a superfield formulation for a generic $\SU(N)$ self-dual field is more complicated and is not solved yet.
However, by introducing extra variables $\varphi_i$, it is always possible to write a {\it component} Lagrangian
(\ref{eq_lagrNA}) (together with the first line in (\ref{eq_action})) corresponding
to the matrix Hamiltonian (\ref{eq_susyham}).

This observation has actually nothing to do with supersymmetry. It boils
down to the following. Consider the eigenvalue problem for a usual Hermitean matrix $H_{jk}$. It can be treated as a Schr\"odinger problem
$\hat H \Psi(\varphi_j) = \lambda \Psi(\varphi_j)$ with the constraint $\hat G \Psi = 0$, where
 \begin{equation}
\label{HGphi}
 \hat H \ =\ \varphi_j H_{jk} \frac {\partial}{\partial \varphi_k} \ , \ \ \ \ \ \ \ \ \ \ \hat G = \varphi_j
\frac {\partial}{\partial \varphi_j} - 1\ .
 \end{equation}
The corresponding Lagrangian
 is
\begin{equation}
\label{Lphi}
L \ =\ i\bar \varphi_j \dot \varphi_j - B(\bar \varphi_j \varphi_j - 1) -  \bar\varphi_j H_{jk} \varphi_k\ ,
\end{equation}
where $\bar\varphi_j=(\varphi_j)^*$.
This easily generalizes to the case where $H$ is an operator depending on a set of canonically conjugated variables
$\{p_\mu, x_\mu \}$. The only difference is that $-H_{jk}$ is now replaced  by the matrix $L_{jk}$ obtained from $H_{jk}$ by the appropriate
Legendre transformation.\footnote{This elementary observation should be well known, for example, in matrix models.
 Surprisingly, we have not found it in such a ``chemically pure'' form in the
literature, but similar constructions were discussed, e.g., in Refs.~\cite{Polych,topkvant}.}

Our initial goal was to find a Lagrangian representation for the Hamiltonian (\ref{eq_susyham}) with matrix-valued $\A_\mu$, $\F_{\mu\nu}$.
The construction just described, with $\varphi_i$ in the fundamental representation of $\SU(N)$, leads
to the $N\times N$ matrix Hamiltonian. The Lagrangian (\ref{Lphi}) coincides in this case with the Lagrangian (\ref{eq_lagrNA})
with the choice $k=1\,$, to which the first line from Eq.~(\ref{eq_action}) is also added.

Obviously, one can describe the Hamiltonians in higher representations of $\SU(N)$ in a similar way, by choosing the number of components $\varphi_i$
equal to the dimension of the representation. We have seen, however, that in the $\SU(2)$ case one can be more economic, introducing
only a couple of dynamic variables $\bar\varphi^\alpha$, but multiplying the term $\sim B$ in the Lagrangian by an arbitrary
integer $k$. This leads to the Hamiltonian in the representation of spin $|k|/2$. Certain $\SU(N)$ representations
(namely, the symmetric products of $|k|$ fundamental or $|k|$ antifundamental representations) can also be attained in this way.

The Lagrangians (\ref{eq_action}), (\ref{eq_lagrNA}) are invariant, up to a total derivative, with respect
to the following infinitesimal ${\cal N}=4$ supersymmetry transformations:
\begin{equation}\label{susytran}
\begin{array}{c}
  x_\mu\rightarrow x_\mu
  +f\epsilon\sigma_\mu\psi
  +f\bar\epsilon\sigma_\mu\bar\psi,
\\[3mm]
  f\psi_{\dot\alpha}\rightarrow f\psi_{\dot\alpha}
  +i\dot x_\mu \left(\bar\epsilon\sigma_\mu\right)_{{\dot\alpha}},
\\[3mm]
  f\bar\psi^{\dot\alpha}\rightarrow f\bar\psi^{\dot\alpha}
  -i\dot x_\mu \left(\sigma^\dagger_\mu\epsilon\right)^{{\dot\alpha}},
\\[3mm]
  \varphi_i\rightarrow \varphi_i
   +i f\left(t^a\varphi \right)_i \A_\mu^a
    \left(\epsilon\sigma_\mu\psi+\bar\epsilon\sigma_\mu\bar\psi\right),
\\[3mm]
  \bar\varphi^i\rightarrow \bar\varphi^i
   -i f\left(\bar\varphi t^a\right)^i \A_\mu^a
    \left(\epsilon\sigma_\mu\psi+\bar\epsilon\sigma_\mu\bar\psi\right).
\end{array}
\end{equation}

Obviously, one can also construct in this way a ${\cal N} =2 $ supersymmetric Lagrangian for the Hamiltonian (\ref{eq_susyham})
with generic (not necessarily self-dual) ${\cal A}_\mu$. A similar construction (but with extra fermionic rather than bosonic variables)
was in fact discussed in Ref.~\cite{Gaume}.
A beauty of the HSS approach explored in this paper is, however, that such extra variables and the constraint
(\ref{svjazk}) are not introduced by hand, but
arise naturally from the manifestly off-shell supersymmetric superfield actions.

Among the directions of further study, it is worthwhile to mention the construction of higher ${\cal N}$ SQM models
with non-Abelian gauge field backgrounds, e.g. ${\cal N}=8$ ones, making use of a nonlinear counterpart of $q^{+}$ \cite{Delduc}
to describe the basic 4-manifold (in this case, the bosonic geometry is not conformally-flat), as well as studying
various supersymmetry-preserving reductions of these models to lower-dimensional target bosonic manifolds
by the gauging procedure of \cite{Delduc}. Actually, the method of the auxiliary ``semi-dynamical'' $({\bf 4,4,0})$
multiplet with the Wess-Zumino type action, which we successfully applied in our construction here, could work with
the equal efficiency for constructing a Lagrangian description of other supersymmetric quantum-mechanics problems involving
the coupling to an external non-Abelian gauge field. Besides the obvious examples
of quantum Hall effect (or Landau problem) in higher dimensions (see e.g. the discussion in \cite{gknty}), we would like to mention
supersymmetric Wilson loop functionals which can be interpreted in terms of a non-Abelian version of Chern-Simons
(super)quantum mechanics \cite{HT}, with the parameter along the loop as an evolution parameter. We hope that
the quantized semi-dynamical variables could provide a new efficient tool to study this class 
of problems.

\section*{Acknowledgments}
E.I. thanks SUBATECH (Nantes) for the warm hospitality extended to him during the course of this work.
He acknowledges a partial support from the RFBR grants 08-02-90490, 09-02-01209 and 09-01-93107.
E.I. and M.K. thank the University of Lyon, where a part of this work was fulfilled, for hospitality.
The authors  are grateful to F.~Delduc and A.~Gorsky for interest in the work and useful discussions. M.K. thanks O.~Driga for encouraging
during this work. The work of M.K. was supported in part by the RFBR grant 07-02-01161 and
a grant for the leading scientific schools NSH-3036.2008.2.

\end{document}